\documentclass{desyproc}
\usepackage{cite}

\begin{document}
\title{Forward-backward multiplicity correlations in pp collisions at high energy in Monte
Carlo model with string fusion}

\author{{\slshape Vladimir Kovalenko, Vladimir Vechernin}\\[1ex]
Saint Petersburg State University, Russia}


\doi 

\maketitle

\begin{abstract}
%
The correlations between multiplicities in two separated rapidity windows, 
is studied in the framework of the Monte Carlo 
model
based on the
picture of string formation in elementary collisions of colour dipoles. The hardness of the elementary collisions is defined by a
transverse size of the interacting dipoles.
The dependencies of the
forward-backward correlation strength on the width and position of the pseudorapidity windows, as well as on transverse momentum
range of observed particles were studied.
It is demonstrated that taking into account of the string fusion effects improves the agreement with the available experimental data. 
\end{abstract}

\section{Introduction}


Long-range correlation studies between observables in two separated rapidity windows
are considered \cite{b_ALICE} as a tool for investigation
of the initial stages of the hadronic and nuclear collisions, 
preceding the creation of 
a hot and dense medium.
Because of the non-perturbative nature of multiparticle production in a soft region, one has to apply the various semiphenomenological approaches, such as 
the model of quark-gluon string formation.
At high energies, due to multiparton interactions, the formation of several pairs of strings becomes possible. The interaction between the strings could be observed as a collective phenomena in pp collisions.

Experimentally, the multiplicity correlation coefficient, defined as 
$b_{\mathrm{corr}}=\frac{\langle n_B n_F \rangle - \langle n_B \rangle \langle n_F \rangle}{\langle n_F^2 \rangle - \langle n_F \rangle_2}$, has been measured in a wide energy
range \cite{ALICE_baldin, ATLAS_bcorr, UA5_bcorr} as a function of pseudorapidity windows width, their position and transverse momentum region.
In the present paper, we study $b_{\mathrm{corr}}$ using the Monte Carlo model, that incorporates string collectivity effects in the form of string fusion \cite{SF}, and compare the results with the data at $\sqrt{s}$ from 200 to 7000 GeV.


\section{Monte Carlo model}

The Monte Carlo model \cite{MC_model} is based on the partonic picture of nucleon interaction. It preserves the energy and angular momentum conservation in the nucleon initial state 
and uses the dipole approach \cite{hardness} for description of elementary partonic collisions.
Multiplicity and transverse momentum are obtained in the approach of colour strings, stretched between projectile and
target partons.
The interaction of strings is realized in accordance with the
string fusion model prescriptions \cite{SF}.
Namely the mean multiplicity $\mu$ and
 the mean transverse momentum $p_{T}$ of the particles produced from a cluster of $k$ overlapping strings are related to those ($\mu_1, {p_T}_1$) from a single string: 
$\mu= \sqrt{k} \mu_{1}, p_{T}=\sqrt[4]{k} {p_T}_{1}.$ 
For realization of the string fusion prescription, we have used the discrete approach, in which a lattice with the cell area equal to the string transverse area $\pi r_{\mathrm{str}}^2$ is introduced. The strings are
thought to be fused if their transverse position centres belong to the same cell.
For the multiplicity from one string (or a cluster of fused strings) we used Poisson
distribution, with Gaussian transverse momentum spectra of produced particles.

However, in order to provide to provide the possibility of a direct comparison with experimental data, 
the correct description of the transverse momentum spectra is required. For this purpose the
MC model \cite{MC_model} has been extended by taking into account the hardness of elementary collision. For this the mechanism similar to the one in DIPSY event generator \cite{hardness}, has been incorporated in our model
with the string fusion.
It was assumed that the hardness an elementary collision is inversely proportional to the
transverse size of the interacting dipoles:
$
{d}_i=|\vec r_1 - \vec r_2 |,  d_i'=|\vec {r_1}' - \vec {r_2}'|.
$
The mean transverse momentum of particles produced by a single string has the contributions from both edges of the string plus the additional constant term $p_0$, corresponding to the intrinsic string transverse momentum:
${{p_{T}}_{1}}^2=\frac{1}{d_i^2}+\frac{1}{{d_i'}^2} + p_0^2$.
Accordingly, in the version with string fusion, the transverse momentum of a cluster of strings: 
$p_T^4 = \sum_{i=1}^k {{p_{T}}_{1_i}}^4,$
where ${{p_{T}}_{1_i}}^2=\frac{1}{d_i^2}+\frac{1}{{d_i'}^2} + p_0^2$.

Parameters of the model are constrained from the data on total
inelastic cross-section and multiplicity\cite{MC_model}. In the present study, for the case with string fusion we have used $r_{\mathrm{str}}\mathrm{=0.2fm}$
(in the case with string fusion). For the intrinsic string transverse momentum we have used ${p_0 = 0.2 \mathrm{ GeV}/c}$, which provides a reasonable description of the transverse momentum distribution in pp collisions at the LHC energies.

\section{Results}
\label{sec:figures}
\begin{figure}[hb]
\centerline{\includegraphics[width=0.9\textwidth]{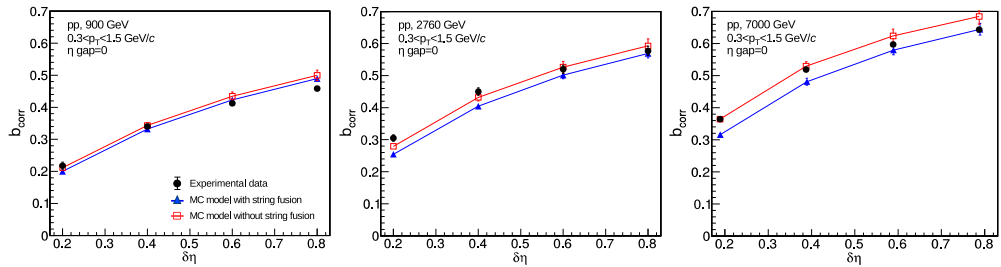}}
\vspace{-0.4cm}
\caption{Correlation coefficient as a function of the pseudorapidity windows width ($\delta \eta$) at midrapidities  ($\eta \ \mathrm{gap} =0$). Lines -- the results of calculation in the model with and without string fusion, points -- the experimental data \cite{ALICE_baldin}.}\label{Fig:MV}
\end{figure}

Figure \ref{Fig:MV} shows the dependence
of the correlation coefficient on the
width of the pseudorapidity windows 
at three energies. The cuts on the transverse
momentum ($\mathrm{0.3}<p_T <\mathrm{1.5 GeV/}c$) applied in MC model calculations, enable direct comparison with the ALICE experimental data \cite{ALICE_baldin}. It was found that the
general trends, like the growth of $b_\mathrm{corr}$ with 
collision energy and width of pseudorapidity windows) are well described by the model.
The role of string fusion raises with $\sqrt{s}$, but using only midrapidity experimental data on multiplicity correlation coefficient at present energies it is hard to distinguish between cases with and without string fusion. Also, it should be noted that at the small gap between rapidity windows there is a contribution of short-range correlation effects in the data (such as the decays of resonances), which are currently not accounted by the model.

\begin{figure}[h]
\centerline{\includegraphics[width=0.35\textwidth]{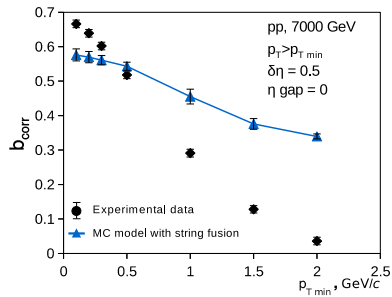}}
\vspace{-0.1cm}
\caption{Correlation coefficient as a function of the lower transverse momentum bound. Line -- the result of calculation in Monte Carlo model with string fusion, points -- the experimental data \cite{ATLAS_bcorr}.}\label{Fig:2}
\vspace{-0.1cm}
\end{figure}

In Fig.~2 the dependence of $b_\mathrm{corr}$ on the transverse momentum region of accounted particles is shown. The correlation coefficient is studied as the function of the lower bound of the $p_T$ interval, and compared to the ATLAS experimental data \cite{ATLAS_bcorr}.
The qualitative agreement of $b_\mathrm{corr}$ with experimental data is found. 
The increase of ${p_T}_\mathrm{min}$ is
accompanied by the decrease
of the multiplicity in the given 
transverse momentum region, which 
restricts the phase space for particle production and the number of ``active'' strings, which leads to the decrease of correlation coefficient.
Numerically, the model calculations overestimate the value of the correlation
coefficient in the hard transverse momentum 
area. It could be an indication
that the direct approach with the soft strings
is applicable at the $p_T$ region below $\sim 1 \mathrm{GeV}/c$, and different
processes (such as jet fragmentation)
begin playing a role at higher $p_T$. 
On the other hand, this dependence is very sensitive
to the shape of the transverse momentum spectra.
 Whereas,
 the description of the $p_T$ spectra in
our approach has an effective character and
does not account  the jet production and other hard phenomena accurately.

Figure~\ref{Fig:3} shows the dependence of $b_\mathrm{corr}$ on the $\eta \mathrm{\ gap}$ at four energies, calculated in the Monte Carlo model, with a comparison to the experimental data \cite{ATLAS_bcorr,UA5_bcorr}. The model calculations do not discriminate pp and $\mathrm{p}\bar{\mathrm{p}}$ scattering.
The model reproduces the growth of the correlation coefficient with collision energy and qualitatively describes the decrease of the correlation coefficient with with increase of the gap between pseudorapidity windows. Note that the short-range effects, such as the resonances decays and jets, which could contribute to the correlation coefficient at small $\eta \mathrm{\ gap}$ are
not accounted by the model.
The results indicate that 
taking into account of the string fusion effects improves the agreement with the data.


\section{Summary and conclusions}
The forward-backward multiplicity correlation strength in pp collisions at high energy is studied in the Monte Carlo model with string
formation and fusion.
The Monte Carlo model reasonably describes the main
features of the behaviour of the correlation coefficient in a wide energy range, such as general growth of the correlation coefficient with collision energy and with
 increase of pseudorapidity window size. The decrease of $b_\mathrm{corr}$ with the increase of the gap between
windows and with increase of the lower $p_\mathrm{T}$ bound  is also qualitatively described. It is found that the version of the model with inclusion of string fusion effects is better supported by the data compared to the case without string fusion.

\section{Acknowledgments}

The authors acknowledge Saint-Petersburg State University for the research grant
11.38.197.2014. V. Kovalenko acknowledges Saint-Petersburg State University for the Special Rector's Scholarship. He is also grateful to the
Dynasty Foundation.


\begin{footnotesize}

\begin{figure}[h]
\centerline{\includegraphics[width=0.9\textwidth]{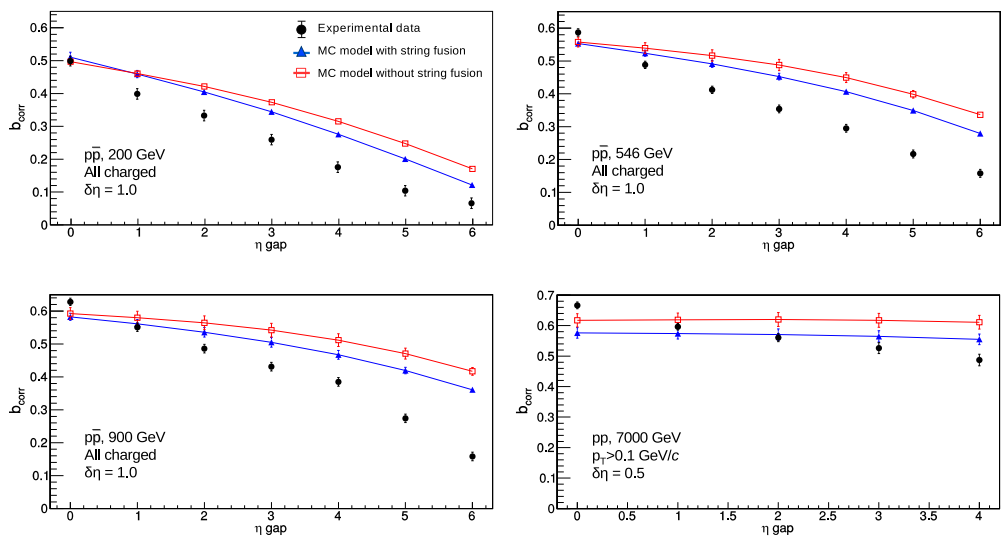}}
\vspace{-0.3cm}
\caption{Correlation coefficient as a function of the pseudorapidity gap. Lines -- calculation in Monte Carlo model with and without string fusion, points -- experimental data \cite{ATLAS_bcorr,UA5_bcorr}.}\label{Fig:3}
\vspace{-0.3cm}
\end{figure}



%

\end{footnotesize}


\end{document}